# The Intermediate Phase in Ternary $Ge_xAs_xSe_{1-2x}$ Glasses

Tao Qu, D.G. Georgiev, P. Boolchand and M. Micoulaut[1]
Department of Electrical, Computer Engineering and Computer Science,
University of Cincinnati, Ohio 45221-0030, USA
[1] Laboratoire de Physique Theorique des Liquides
Universite Pierre et Marie Curie, Boite 121, 4 Place Jussieu
75252 Paris, Cedex 05, France

## ABSTRACT

Melt-quenched $As_xGe_xSe_{1-2x}$ glasses over the composition range, $0 < x < 0.26$, are examined in Raman scattering, T-modulated Differential Scanning Calorimetry (MDSC), and $^{119}$Sn Mossbauer spectroscopy measurements. The non-reversing enthalpy near $T_g$, $\Delta H_{nr}(x)$, accessed from MDSC shows a global minimum ($\sim 0$) in the $x_c(1) = 0.09 < x < x_c(2) = 0.16$ range, and increases by an order of magnitude both at $x < x_c(1)$ and at $x > x_c(2)$. Raman mode frequency of corner-sharing $Ge(Se_{1/2})_4$ tetrahedra studied as a function of x, also shows *three* distinct regimes (or power-laws, p) that coincide with $\Delta H_{nr}(x)$ trends. These regimes are identified with *mechanically floppy* ($x < x_c(1)$), *intermediate* ($x_c(1) < x < x_c(2)$), and *stressed-rigid* ($x > x_c(2)$) phases. The Raman elasticity power-law in the *intermediate phase*, $p_1 = 1.04(3)$, and in the *stressed rigid* phase, $p_2 = 1.52(5)$, suggest effective dimensionalities of $d = 2$ and 3 respectively.

## STRUCTURE BASED CLASSIFICATION OF GLASSES

In the early 1980s, Phillips [1] and independently Thorpe [2] suggested that a network of polymeric chains (weakly crosslinked) will spontaneously stiffen or become rigid when chain-cross-linking acquires a threshold value. In covalent systems, one usually expresses the degree of cross-linking in terms of a mean coordination number, *r*. The *floppy to stressed rigid* elastic phase transition was predicted to occur near *r* = 2.40. However, recent Raman scattering [3-5] and independently T-modulated DSC experiments [3-8] on chalcogenide glasses have shown that there are in fact *two* (rigidity transitions) and not *one* transition as predicted by mean-field constraint theory. Thus, the onset of stressed rigidity in disordered systems display a far richer structure [9] than previously believed. Specifically, the frequency of corner-sharing $Ge(Se_{1/2})_4$ tetrahedra in $Ge_xSe_{1-x}$ glasses [5] display a kink (change in slope) near x = 0.20 (or *r* = 2.40 ; *transition 1*), and a discontinuous jump between x = 0.25 and 0.26 (*r* = 2.55; *transition 2*) followed by a distinct power-law at higher x. Parallel results are now available in binary Si-Se [3], As-Se[6] and P-Se [7] glasses. These results provide evidence for the opening of *intermediate phases* [9-12] between *floppy* and *stressed- rigid* ones in disordered networks.

The physical picture of elastic phase transitions in network glasses above suggests that one can generically classify these disordered networks into *three* distinct phases, ***floppy-***

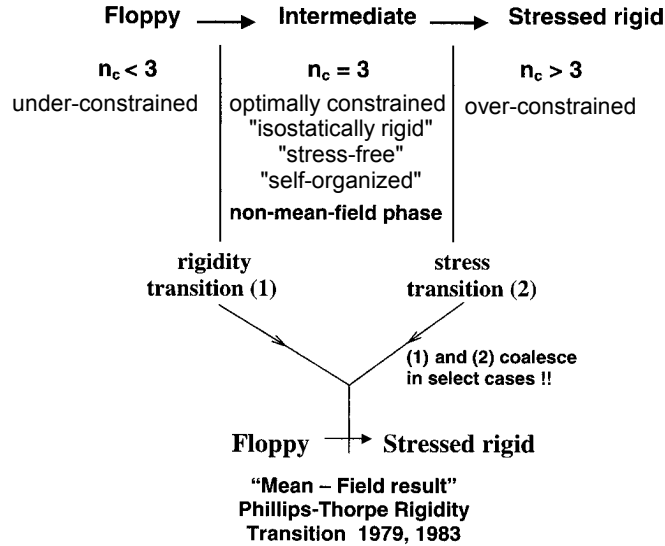

**Figure 1.** Self-organization of disordered networks leads to opening of intermediate phases between floppy and stressed rigid phases.

*intermediate-stressed rigid*. Weakly crosslinked networks in which the count of Lagrangian bonding constraints per atom, $n_c$, is less than 3 belong to *floppy phases*. Optimally crosslinked networks in which $n_c$ is near 3, are *isostatically rigid* and serve to define *intermediate phases*. On the other hand, strongly crosslinked networks in which the count of $n_c$ is greater than 3 belong to *stressed rigid phases* (Fig. 1). Here the count of 3 comes from the degrees of freedom associated with an atom in a 3d network.

## **Phillips-Thorpe Rigidity Transition**

In select systems the *rigidity transition 1* and *stress transition 2* can coalesce, and one observes a *solitary* phase transition as first predicted by Phillips [1] and Thorpe [2] using mean-field constraint theory. Two rather striking examples of the behavior have now been documented and include, (a) a light-induced modification of rigidity transitions [5] in binary Ge-Se glasses and (b) a sharp *floppy* to *stress rigid* transition in ternary Ge-S-I glasses [13]. In binary $Ge_xSe_{1-x}$ glasses, macro-Raman measurements have shown the intermediate phase to extend in the $0.20 < x < 0.25$ region. In macro-Raman measurements the exciting laser beam is brought to a loose focus typically 50 μm or more. On the other hand, in micro-Raman measurements that utilize a microscope attachment, the exciting laser beam is usually brought to a tight focus (1 μm) with the consequence that the flux of exciting radiation is usually three orders of magnitude higher than in macro-Raman measurements. A high photon flux of sub-band gap radiation can lead to rapid switching of bonds and photomelt the backbone of these Ge-Se glasses. A consequence of the underlying photo-structural effect is the loss of self-organization, and results in a sharp and first-order rigidity transition near $x = 0.225$ as reported by Feng et al. [4]. The interpretation of the micro-Raman results became clear only once results of macro-Raman measurements became available more recently [5].

The second illustrative example of a *sharp* rigidity transition [8] is that of ternary $Ge_{1/4}S_{3/4-y}I_y$ glasses. Here iodine for sulfur chemical alloying lowers the global connectivity of the backbone, and a *floppy* to *stressed-rigid* transition is predicted by mean-field theory to occur at $y = 1/6$. Thermal measurements on the ternary glasses are in surprising accord with the mean-field prediction. Why should this be the case ? The answer appears to be that iodine replacement in the ternary alloy glasses apparently scissions the backbone *stochastically* resulting in loss of n-membered rings, with n > 4, where isostatic rigidity is thought to be nucleated [13]. It is because of the absence of self-organization that the floppy to stressed rigid transition is observed to occur exactly where mean-field theory predicts it. A closely similar result of a sharp floppy to rigid transition has now been observed in corresponding selenide glasses by Wang et al. [14]. Each selenium atom like a sulfur atom contributes two constraints per atom as bridging sites coupling tetrahedral units. Thus, constraint counting algorithms do not distinguish between the two chalcogens. These observations underscore that the origin of these physical effects undoubtedly results from elastic phase transitions.

## The special case of the $Ge_xAs_xSe_{1-2x}$ ternary glass system

Figure 2 shows a plot of $T_g$ in indicated binary and ternary glasses as a function of mean coordination number *r*. We notice that in both binary glasses $T_g$ shows a global maximum near the chemical thresholds, while no such maximum is observed in the titled ternary. The threshold behavior in $T_g$ near *r* = 2.4 in the As-Se, and near *r* = 2.67 in the Ge-Se, we have suggested [15], constitutes evidence for demixing of the cation-bearing homopolar bonds (As-As, Ge-Ge) nucleated near these chemical thresholds from the glass backbone. Such nanoscale phase separation is clearly absent in the present ternary wherein $T_g$s continue to monotonically increase even when the cation concentration x exceeds the chemical threshold, $x_t = 0.182$, or *r* = 2.55. The glass transition temperature $T_g$ intimately reflects global connectivity of networks as shown by stochastic agglomeration theory [16]. The cation-bearing homopolar bonds apparently form part of the network backbone in the present glasses for $T_g$ to increase monotonically. This is an attractive feature of the ternary glass system for probing rigidity transitions. Presence of nanoscale phase separation effects in glasses usually produces pronounced changes in glass physical properties that can mask the more subtle connectivity related effects contributing to rigidity transitions.

In the present work we have now examined the nature of the rigidity transitions [8] in ternary $As_xGe_xSe_{1-2x}$ glasses using Raman scattering. The evidence obtained confirms the existence of *three* elastic phases with a rather wide *intermediate phase*. The feature has permitted a reliable measurement of the elastic power-law in that phase. Furthermore, the striking jump in mode frequency shift [11] of corner-sharing $Ge(Se_{1/2})_4$ tetrahedra between the *intermediate* and the *stressed-rigid phase* suggests that the *stress transition 2* is **first-order** (Fig. 1) in character here.

## Chemical threshold ($x_t$) in $As_xGe_xSe_{1-2x}$ ternary

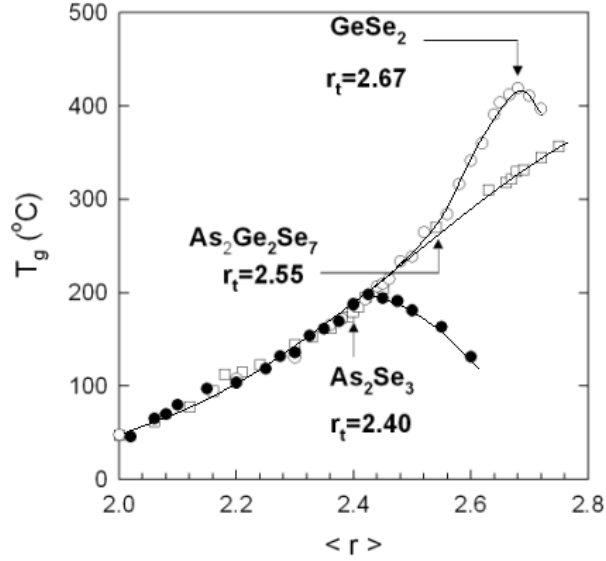

**Figure 2**. Variations in $T_g(r)$ in As-Se (●), Ge-Se (○) and $Ge_xAs_xSe_{1-2x}$ (□) glasses as a function of mean coordination number $r$. Note the existence of global maxima in $T_g$ near chemical thresholds, $r_t$, in the binary glasses, but not in the ternary one. The chemical threshold in the ternary occurs at $x_t = 2/11$, or mean $r_t = 2.55$.

The chemical threshold in the present ternary represents the chemical composition that leaves no free Se in the network. At $x > x_t$, one thus expects homopolar ( As-As, (Ge-Ge) bonds to appear in the network. On stoichiometric grounds one can write,

$$As_xGe_xSe_{1-2x} = [(5x/2)(As_{2/5}Se_{3/5})][(3x)(Ge_{1/3}Se_{2/3})][\ Se_{1-2x-3x/2-2x}] \qquad (1)$$

The chemical threshold would be manifest when the free Se fraction vanishes, i.e.,

$$1-2x_t - 3x_t/2 - 2x_t = 0$$

or $\qquad\qquad x_t = 2/11 = 0.1818 \qquad (2)$

Formally, the mean coordination number, $r_t$, corresponding to the threshold becomes

$$r_t = 2 + 3x_t = 28/11 = 2.55 \qquad (3)$$

We note the absence of a threshold behavior in $T_g$ near $r_t$ (Fig. 2) underscoring the absence of nanoscale phase separation effects in the present ternary glasses.

**EXPERIMENTAL**

**Temperature Modulated DSC**

The starting materials to synthesize the bulk $As_xGe_xSe_{1-2x}$ glasses in the $x < x_t$ range consisted of 99.999 % $As_2Se_3$, Ge and Se from Cerac Inc. Glass compositions at $x > x_t$, required the use of elemental As in addition, and the starting material required special handling. Details of the synthesis appear in ref [7,8]. Once water quenched, the glasses were allowed to relax at room temperature for 3 weeks or longer prior to initiating any physical measurements.

Traditionally $T_g$s are established [17] in differential scanning calorimetry (DSC) measurements, as the inflexion point of an endotherm observed when a glass sample is heated at a certain scan rate (typically 20°C/min), in which T is programmed to increase linearly with time (t). The introduction of temperature modulated DSC (MDSC) has significantly enhanced understanding the nature of glass transition [6-8,18,19]. MDSC permits to separate ergodic from non-ergodic processes that contribute to softening of a glass. This is achieved by programming a sinusoidal T-modulation over the linear T-ramp. The fraction of the heat flow that tracks the T-modulation is obtained by a Fourier analysis of the total heat-flow. These heat-flow terms are illustrated in Fig. 3 for a glass sample at x = 0.08 examined using a model 2920 MDSC from TA instruments Inc. One can see that the reversing heat flow signal (Fig. 3) shows a characteristic step that one normally ascribes to a glass transition. An advantage of MDSC is that the reversing heat flow signal *always* sits on a flat baseline independent of the equipment baseline, and one can extract $T_g$ as the inflexion point of the step transformation. The difference signal between the *total-* and the *reversing- heat flow* is termed as the *non-reversing heat flow*. In programming our MDSC scans, we scan up in T across the glass transition temperature and then back down in T, to obtain the mean-value of $T_g$. The $T_g$ so obtained, is not only independent of scan rates but also of sample thermal history [6]. The latter is contained in the non-reversing heat flow signal that usually shows a Gaussian-like profile as a precursor to the reversing heat-flow signal (Fig. 3). The integrated area under the non-reversing heat flow signal (shaded area in Fig. 3) is the non-reversing relaxation enthalpy, $\Delta H_{nr}(x)$. The advantage of scanning up and down in T is that one can correct the $\Delta H_{nr}(x)$ term for the small shift of the reversing heat flow signal because of the small but finite scan rate used. The frequency corrected $\Delta H_{nr}(x)$ term is plotted in Fig 4(b). One finds that in the $0.09 < x < 0.16$ region the $\Delta H_{nr}(x)$ term vanishes, and increases to 0.5 to 1.0 calories/gm at $x < 0.09$ and at $x > 0.16$.

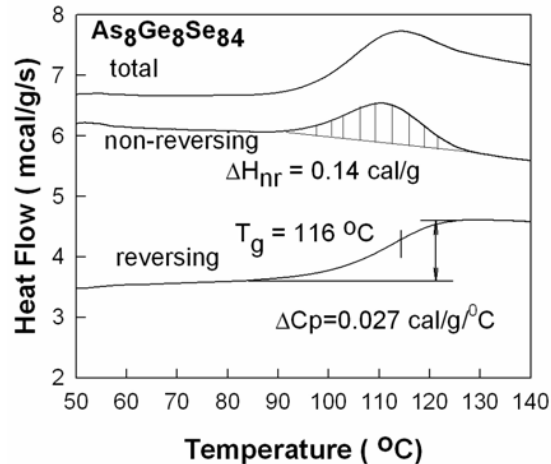

**Figure 3**. MDSC scan of titled ternary glass taken at a scan rate of 3°C/min and a modulation rate of 1°C/100sec, showing the total heat flow, deconvoluted in terms of non-reversing and reversing heat flows. The shaded area gives the non-reversing enthalpy $\Delta H_{nr}$, while the inflection point of the reversing heat flow gives $T_g$.

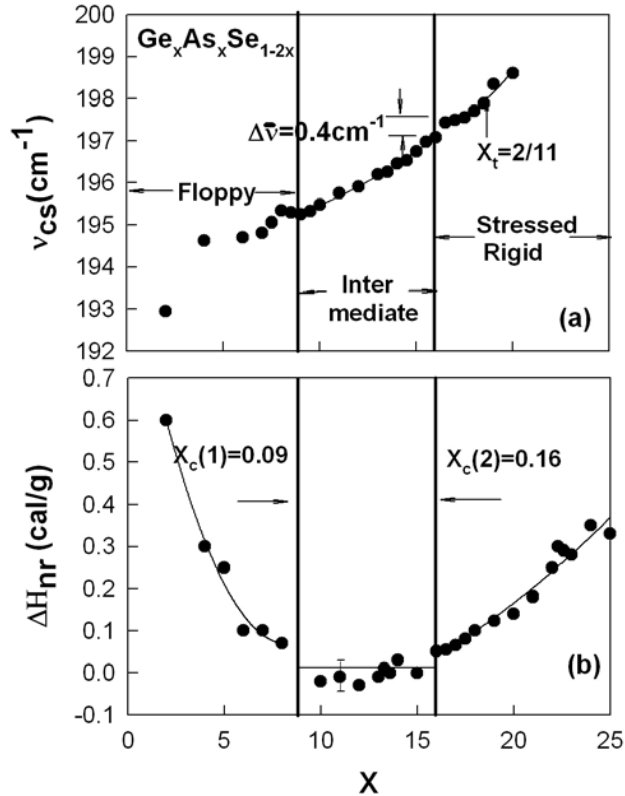

**Figure 4.** Compositional trends **(a)** Raman mode frequency of corner-sharing $Ge(Se_{1/2})_4$ tetrahedra **(b)** non-reversing enthalpy [8], $\Delta H_{nr}(x)$, in the present glasses displaying three regimes, floppy ( $x < 0.09$), intermediate( $0.09 < x < 0.16$) and stressed rigid ( $x > 0.16$) that coincide with each other. In the intermediate phase the $\Delta H_{nr}(x)$ term vanishes.

These $\Delta H_{nr}$ results highlight the very special nature of glass compositions in the $0.09 < x < 0.16$ composition range. Glass transitions for these compositions are almost completely thermally reversing in character. Such glass samples and their corresponding melts are configurationally very similar to each other. And it is therefore not surprising that such melts apparently form glasses even when cooled very slowly. We will comment on these glasses further in relation to Raman scattering results later.

**Mossbauer Spectroscopy**

250 days $^{119m}Sn$ parent in a $CaSnO_3$ matrix was used as an emitter of 23.8 keV γ-rays to record Mossbauer spectra [20] of glasses at 4.2 K using a standard constant acceleration drive and a liquid helium exchange gas dewar. Glass samples were doped with ½ weight percent of isotopically enriched $^{119}Sn$ in its elemental form. The doping was performed by alloying the base glass with the dopant at 950ºC for 24 hours in evacuated quartz tubing. The melts were equilibrated at 50ºC above the liquidus before a water quench. Figure 5 reproduces spectra at select glass compositions near the chemical threshold $x_t = 0.18$. One observes (Fig. 5) two sites in the lineshapes, a singlet (site A) and a doublet (site B). The integrated intensity of site B, $I_B/[I_A + I_B]$, is found to display a threshold behavior as shown in Fig. 6. Site A (isomer shift $\delta_A = 1.64(1)$ mm/s; quadrupole splitting $\Delta_A = 0.41(1)$ mm/s) is identified [20,21] with Sn replacing Ge in a tetrahedrally coordinated $Ge(Se_{1/2})_4$ units. Site B ($\delta_B = 3.36(2)$ mm/s; $\Delta_B = 1.40(2)$ mm/s) is identified with Sn replacing Ge in $Ge_2Se_3$ units. The latter reflect Ge-Ge bonds in the network, and as expected these nucleate near the chemical threshold ($x_t$) (Fig.6).

**Raman scattering**

Raman scattering measurents were undertaken with both a dispersive system and an FT-Raman system. The dispersive system used a T64000 triple monochromator instrument from Jobin Yvon Inc. in a macro configuration with samples contained in fused quartz tubes [5]. The spectra were taken at 4 cm$^{-1}$ resolution. The FT-Raman measurements used a Nicolet FTIR 870 bench coupled to a Raman module [22]. The FT-Raman measurements utilized a Nd-YAG laser with 1.06 micron (1.24 eV) excitation and gave two orders of magnitude higher signal to noise than the dispersive system. The FT-Raman spectra were recorded at 1 cm$^{-1}$ resolution with a laser power of 90 milliwatts on a spot size of 1mm$^2$. In this work we report on the FT-Raman results obtained under these conditions. Figure 7 shows observed Raman lineshapes at indicated glass compositions. Alloying As and Ge in a Se glass, shows modes of $As(Se_{1/2})_3$ pyramids and $Ge(Se_{1/2})_4$ tetrahedra to emerge at the expense of modes of $Se_n$ chains. The observed lineshapes were fit to a superposition of Gaussians to extract mode frequency shift of CS tetrahedra ($\nu_{CS}(x)$ ) as a function of x ( Fig. 4a). One can observe three distinct regimes of $\nu_{CS}(x)$; regime 1 in which $\nu_{CS}(x)$ is independent of x at x < $x_c(1)$ = 0.09, a second regime wherein $\nu_{CS}(x)$ displays a linear power-law, $p_1$ = 1.04(3) in the 0.09 < x < 0.16 range, and finally a third regime wherein $\nu_{CS}(x)$ shows a jump near x = 0.16 and displays a power-law, $p_2$ = 1.52(5). These power-laws were extracted by fitting the observed mode frequency variation to the following relation [4,23],

$$\nu_{CS}^2(x) - \nu_{CS}^2(x_c(1)) = A[\, x - x_c(1)]^p \qquad (4)$$

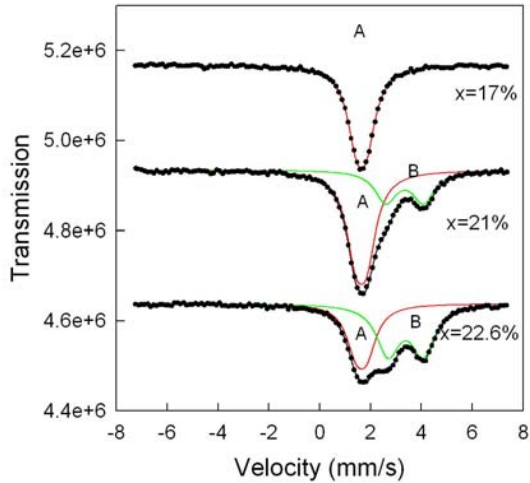 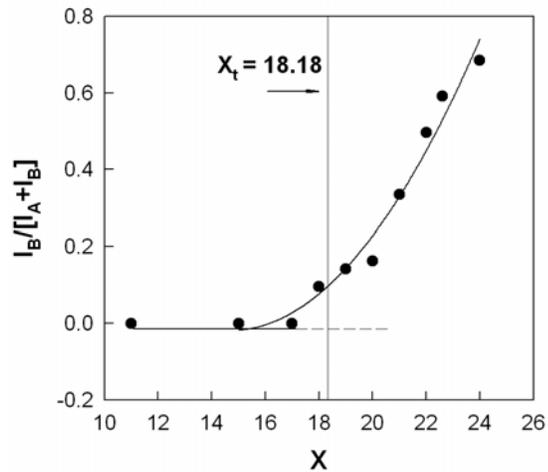

**Figure 5.** Mossbauer spectra of $Ge_xAs_xSe_{1-2x}$ glasses at indicated cation concentrations showing evolution of the B-site doublet near the chemical threshold $x_t = 0.18$. The ratio $I_B/I$ as a function of x appears in Fig.6.

**Figure 6.** Compositional trends in the site integrated intensity ratio $I_B/[I_A+I_B]$ showing growth of B-site doublet near $x_t$. B-sites represent signature of Ge-Ge bonds formed in the network.

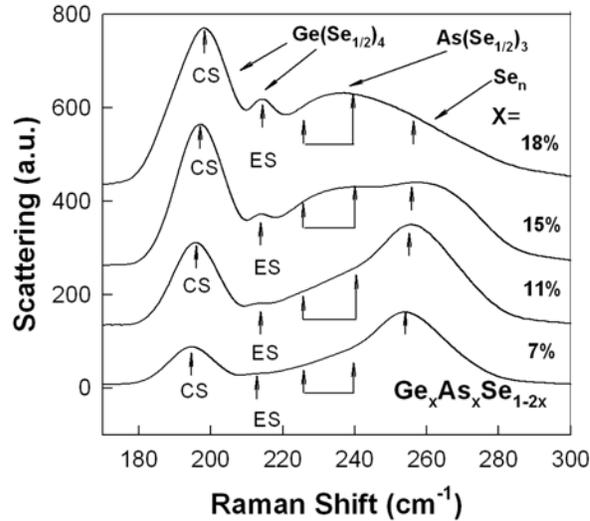

**Figure 7.** Observed FT-Raman lineshapes in glasses showing growth in scattering strength of vibrational modes of corner-sharing ( CS:near 200 cm$^{-1}$) and edge-sharing (ES:217cm$^{-1}$) Ge(Se$_{1/2}$)$_4$ tetrahedra at the expense of Se$_n$ chain mode (250 cm$^{-1}$) with increasing x. The two principal modes of As(Se$_{1/2}$)$_3$ pyramidal units sit at 225 and 243 cm$^{-1}$.

Here $\nu_{CS}(x_c(1)$ represents the observed Raman frequency at the threshold composition $x_c(1)$. In the intermediate phase, the power-law $p = p_1$ was obtained by plotting log $[\nu_{CS}^2(x)- \nu_{CS}^2(x_c(1))]$ as a function of log (x - $x_c(1)$), and the slope of the observed line (Fig. 8a) gave $p_1$= 1.04(3). A parallel plot for glass compositions in the 0.16 < x < 0.20 was undertaken, and yielded a power-law (Fig. 8b) for the stressed rigid phase of $p_2$ =. 1.52(5). In establishing the elastic power-law $p_2$ we choose not to consider glass compositions at x > 0.20 largely because the Ge-Ge mode located near 180 cm$^{-1}$ begins to emerge and overlap with the symmetric stretch of CS mode, making an unambiguous deconvolution of the line shape difficult.

## DISCUSSION

### Molecular Structure of As-Ge-Se glasses

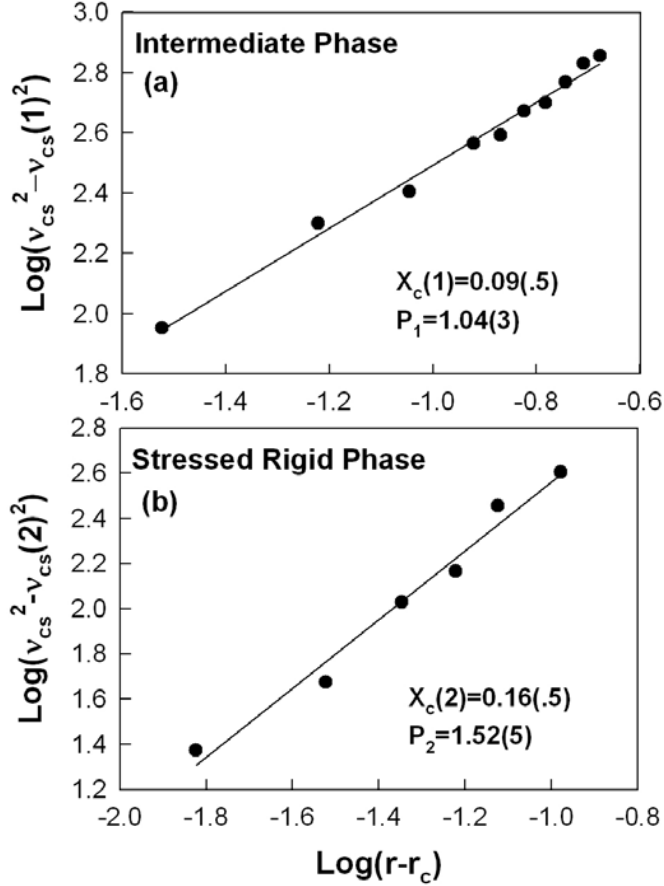

**Figure 8.** Power-law of optical elasticity in the intermediate phase (a) $p_1 = 1.04(3)$ and in the stressed rigid phase (b) $p_2 = 1.52(5)$ deduced from the observed Raman mode frequency shift of CS tetrahedral units examined as a function of glass composition x.

The extensive glass forming tendency in the Ge-As-Se ternary [24], and the three distinct elastic phases are illustrated in Fig. 9. The intermediate phase in the As-Ge-Se ternary is shown by the hashed region between the floppy (red) and the stressed rigid phases. It was constructed by combining the present results with those on binary As-Se and Ge-Se glasses. The intermediate phase straddles the $r = 2.40$ broken blue line. Glasses in the stressed rigid phase are generally nanoscale phase separated (grey shading) along binary compositions [15]. However, ternary glasses containing equal concentrations of As and Ge in a Se glass progressively cross-link $Se_n$ chains forming tetrahedral $Ge(Se_{1/2})_4$, pyramidal $As(Se_{1/2})_3$ and quasi-tetrahedral $Se=As(Se_{1/2})_3$ units. The increased connectivity of the network is reflected in the glass transition temperature progressively increasing (fig. 2). Using stochastic agglomeration theory, we can quantitatively account [8] for the observed increase in $T_g$ at low x (< 0.10). At higher x these building blocks begin to connect with each other, and the agglomeration process ceases to be stochastic as certain chemically preferred configurations emerge. Inelastic neutron scattering measurements have shown [25] that floppy modes (at 5 meV) of $Se_n$ chain fragments are progressively erased as the network is increasingly crosslinked by alloying As and Ge.

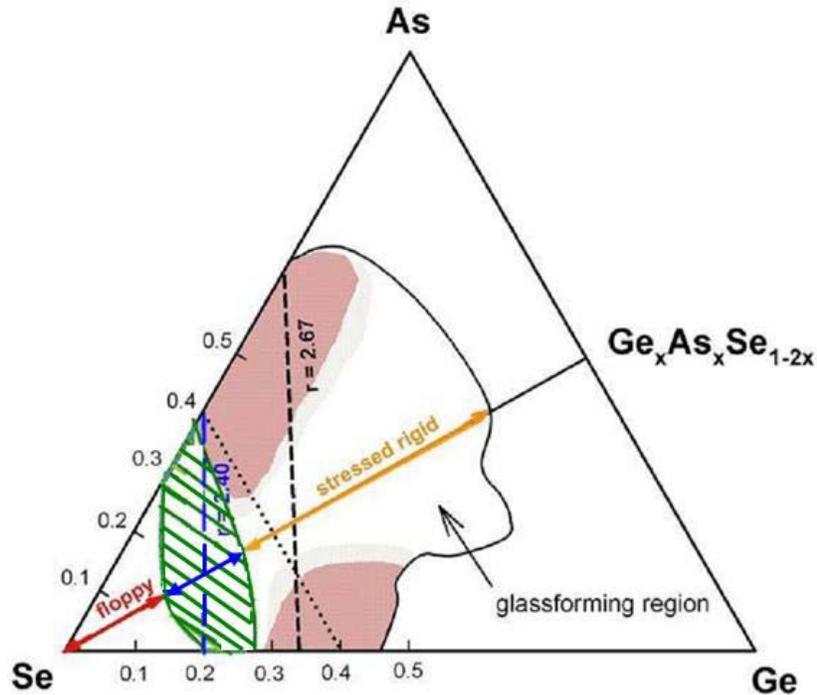

**Figure 9.** The intermediate phase in the As-Ge-Se ternary is shown by the hashed region between the floppy and the stressed rigid phases. It straddles the r = 2.40 broken blue line. Nanoscale phase separated alloys are shown by grey shading.

The local and medium range structures prevailing in the intermediate phase largely are made up of GeSe$_4$ tetrahedra ($r$ = 2.40), As(Se$_{1/2}$)$_3$ pyramids ($r$ = 2.40) and Se=As(Se$_{1/2}$)$_3$ quasi-tetrahedral units ($r$ = 2.28) [26]. Constraint counting algorithms show that these building blocks are isostatically rigid [8]. In other words, a count of bond-stretching and bond-bending constraints per atom, $n_c$, for each of these units reveals a value of exactly 3. One of the reasons the intermediate phase spans such a wide compositional width in the present ternary is that the mean coordination number of these units varies in the 2.28 < $r$ < 2.40 range.

**Elastic Power-law in the Stressed Rigid Phase**

Numerical experiments on depleted amorphous Si three-dimensional networks stabilized by nearest neighbor central (bond-stretching) and non-central (bond-bending forces) have shown that the shear, longitudinal and transverse elastic constants as a function of $r$ show [23,27] a robust power-law with p = 1.5 for *stressed rigid* glasses. The numerical experiments invoke just the type of interactions that Raman vibrational modes probe in a disordered network. The power-law of $p_2$ = 1.52(5) obtained from our Raman measurements on the present ternary glasses at x > 0.16 is in excellent agreement with the numerical experiments above, and serve to demonstrate ( Fig. 8a) that such glasses are *stressed rigid*. Parallel results have been observed from Raman measurements on binary Ge-Se [5] and in Si-Se glasses [3].

Numerical simulations [13] have also shown that stressed rigidity is nucleated in n-membered rings when n < 6. In general, if networks do not possess small sized rings where stressed rigidity can nucleate, onset of stressed rigidity would be catastrophic, i.e. first order. The jump in the corner-sharing mode frequency of $Ge(Se_{1/2})_4$ tetrahedra, $\nu_{CS}$, near the stress transition of $x_c(2) = 0.16$ of $\Delta\nu_{CS} = 0.4$ cm$^{-1}$ (Fig. 4a) is suggestive that the underlying phase transition is *first-order* in character.

## **Intermediate Phase in the $As_xGe_xSe_{1-2x}$ ternary**

The wide compositional width of the intermediate phase and the clearly resolved nature of the CS mode in Raman spectra of these glasses have permitted a reliable measurement of the elastic-power-law in the intermediate phase. The power-law obtained for the intermediate phase (Fig 8b) yields $p_1 = 1.04(3)$. On general grounds [28], one expects the change in stiffness of coherently scattered regions to vary as $(\Delta r)^{d/2}$, where d is the dimensionality of the network. A power-law $p_1 = 1$ could imply, thus, an effective dimensionality of $d = 2$ for the intermediate phase. For stressed-rigid phase, the measured value of $p_2 = 1.5$, would then suggest an effective dimensionality of $d = 3$.
A new approach to understanding the compositional width of self-organization in glassy systems is to use constraint counting algorithms with size-increasing cluster combinatorics [12]. The method was recently used to understand the observed intermediate phases in Si-Se and Ge-Se glasses. The advantage of the method is that one starts with the appropriate local structures that form the elements of the intermediate phase. The medium range structures are inferred by agglomeration of these local structures to infer at what point stressed rigidity becomes manifest.

## CONCLUSIONS

Temperature modulated DSC and Raman scattering measurements on $Ge_xAs_xSe_{1-2x}$ bulk glasses in the $0 < x < 0.26$ range reveal the existence of *three* distinct elastic phases. In the $0 < x < 0.09$ range, glasses are *floppy* possessing increasingly larger non-reversing enthalpy $\Delta H_{nr}$ as x decreases starting from 0.09. In the $0.09 < x < 0.16$ range, glasses possess a vanishing $\Delta H_{nr}$ term and a Raman optical elasticity that varies as a function of mean coordination, r, as a power-law with a power, $p_1 = 1.04(3)$. The behavior is identified with the *intermediate phase* of the present ternary. The phase is thought to be *self-organized* consisting of three distinct optimally coordinated building blocks including $As(Se_{1/2})_3$ pyramidal units, $Se=As(Se_{1/2})_4$ quasi-tetrahedral units and $GeSe_4$ tetrahedral units. In the $0.16 < x < 0.24$ range, the $\Delta H_{nr}$ term increases with x qualitatively, and the Raman optical elasticity is characterized by a power-law, $p_2 = 1.52(4)$. The behavior is identified with the *stressed-rigid* phase of the glasses. The small but finite jump in Raman mode frequency of Ge-centered corner-sharing tetrahedral units of 0.4 cm$^{-1}$ near $x = 0.16$ suggests that the *stress-transition 2*, is *first order* in character. The measured elasticity power-laws in the *intermediate-* and *stressed-rigid* phases suggest that the effective network dimensionalities $d = 2$ and 3 respectively.

## ACKNOWLEDGEMENT

It is a pleasure to thank J.C. Phillips for correspondence during the course of this work. The FT-Raman measurements were made possible by B. Zuk and M. Bradley of Thermo-Nicolet Inc. The work at University of Cincinnati is supported by NSF grant DMR-01-01808. LPTL is Unite Mixte de Recherche associee au CNRS n. 7600.